\documentclass[preprint,5p]{elsarticle}
\usepackage[latin1]{inputenc}
\usepackage{amsmath}
\usepackage{mathptmx}
\usepackage{graphicx}

\begin{document}

\title{On the numerical simulation of Kerr frequency combs using coupled mode equations}

\author[TH]{T. Hansson}
\ead{tobias.hansson@ing.unibs.it}
\author[DM]{D. Modotto}
\author[SW]{S. Wabnitz}
\address{Dipartimento di Ingegneria dell'Informazione, Universit\`a di Brescia, via Branze 38, 25123 Brescia, Italy}

\begin{abstract}
It is demonstrated that Kerr frequency comb generation described by coupled mode equations can be numerically simulated using Fast Fourier Transform methods. This allows broadband frequency combs spanning a full octave to be efficiently simulated using standard algorithms, resulting in orders of magnitude improvements in the computation time.
\end{abstract}

\begin{keyword}
Kerr frequency combs, coupled mode equations, numerical simulation, split-step method, Fast Fourier Transform
\end{keyword}

\maketitle

\section{Introduction}

Kerr frequency comb generation has recently attracted much interest, due to its many potential applications in e.g. spectroscopy, frequency metrology, sensing and arbitrary waveform generation \cite{Kippenberg3}. There are currently two main theoretical formalisms that are being used to model and describe frequency comb generation in microresonator devices. These are the formalisms based on the coupled mode equations \cite{Matsko2,Chembo1}, and the driven and damped nonlinear Schr\"odinger (NLS) equation \cite{Matsko1,Hansson}, the latter being also known as the Lugiato-Lefever equation (LLE) \cite{Erkintalo2,Chembo2}. The driven and damped NLS model has been used since the early '90s for describing nonlinear dispersive fiber-ring cavities \cite{HTW}, but has only recently been adopted as a model for microresonators. The LLE has been widely hailed for allowing significant, orders of magnitude, speedups to the numerical simulation of Kerr frequency combs when compared with simulations using coupled mode equations \cite{Erkintalo2,Chembo2,Gaeta}, thereby permitting simulations of ultra broadband combs such as octave spanning frequency combs consisting of hundreds or thousands of resonant modes.

However, both formalisms have recently been established to describe the same underlying physical phenomena, and can be converted into each other under suitable approximations, see \cite{Chembo2}. Indeed, both formalisms provide an equivalent description of the nonlinear four-wave-mixing process due to the Kerr nonlinearity. There should therefore be no fundamental reason for one method to be substantially faster, with regards to computation, than the other.

In this letter we will show that this is in fact precisely the case. It has previously been demonstrated that the coupled mode equations can be seen as a Fourier expansion of the driven and damped nonlinear Schr\"odinger equation \cite{Hansson}. It is therefore possible to cast the coupled mode equations into a form where Fast Fourier Transforms (FFTs) can be used to speed up calculations. This allows for huge speed improvements, which are analogous to the improvements obtained when using FFTs instead of matrix computations for calculating the discrete Fourier transform, thus potentially shortening computation times by several orders of magnitude. This observation also opens up the possibility of using coupled mode methods for numerical simulations on a larger scale, which is particularly relevant for the simulation of octave spanning frequency combs, consisting of thousands of resonant modes. These octave spanning combs are of great interest from an application point of view since they can be self-referenced to provide the absolute frequency of the comb modes.

We will first give a short overview of how the coupled mode equations are being used in describing Kerr frequency comb generation, and the source of the so far widely perceived difficulty for the associated numerical computations. We will then show how a solution to this problem may be found by reformulating the coupled mode equations in such a way that FFTs can be used to calculate the nonlinear contribution. We will discuss the huge improvement which can be achieved by doing so: as a result, the computation time involved in either the frequency-domain (i.e., the coupled mode equations) or the time domain (i.e., the LLE) approach is exactly of the same order. Set aside computation time, we thus conclude this work by outlining which may be the particular remaining benefits and drawbacks of the different formalisms for studying microresonator frequency combs.

\section{Numerical simulation using coupled mode equations}

The coupled mode equations provide a modal expansion approach for describing optical frequency comb generation in whispering-gallery-mode resonators and microresonator cavities \cite{Matsko2,Chembo1,Kippenberg2}. The coupled mode equations describe the slow evolution of each discrete frequency component of the resonant comb spectrum, with each comb mode $\mu$ modeled by an ordinary differential equation, viz.
\begin{align}
    & \frac{\partial A_\mu}{\partial \tau} = -\frac{\kappa_\mu}{2}A_\mu + \delta_{\mu}\sqrt{\kappa_{ext}}A_{in}e^{i\sigma\tau} + \nonumber\\ & ig_0\sum_{\alpha,\beta,\gamma}A_{\alpha}A^*_{\beta}A_{\gamma}e^{i(\omega_{\alpha}-\omega_{\beta}+\omega_{\gamma}-\omega_{\mu})\tau},
\end{align}
where the first term on the right hand side gives the complete losses $\kappa_\mu$ at frequency $\omega_\mu$ as a sum of both intrinsic and external $\kappa_{ext}$ losses, while the last term gives the contributions to each frequency due to the four-wave-mixing process. The nonlinear coefficient $g_0$ is for simplicity assumed to be a constant, although this is not a necessary restriction, see \cite{Chembo1}. At the central frequency $\omega_0$ there is an additional term which provides the external driving $A_{in}$, with corresponding cavity detuning $\sigma = \omega_{in} - \omega_0$. A single rate equation can be used to model each mode, since it is assumed that the amplitude of each frequency component is evolving slowly on a separate time-scale from the fast time variation corresponding to the temporal profile of the mode.

Simulations of Kerr frequency comb generation using coupled mode equations have so far been widely perceived in the literature as being computationally demanding due to the large number of contributions from the nonlinear four-wave-mixing term, when many modes are present in the cavity. For example, an analysis of the computational complexity was presented in \cite{Chembo1}, which showed that the number of contributions to each equation grows quadratically with the number of modes ($N$) that are considered in the simulation. If numerical methods relying on e.g. loops and conditionals or direct matrix calculations are used, then the computation time will scale cubically to highest order. Although the exact algorithms used by different research groups have not been published, it is obvious from the computation times reported in the literature that quite inefficient algorithms have been used so far for calculating the nonlinear contribution to each frequency. Simulation times of several days have been reported \cite{Chembo1,Kippenberg1}, despite the fact that only a few hundred modes are included in the simulations. It is obviously impractical to simulate combs containing thousands of resonant modes using such numerical codes, since an eightfold increase in computation time is required for each doubling of the number of modes.

However, the cubic scaling property of the computation time is not inherent to the underlying system of coupled mode equations, but it merely depends on the choice of algorithm which is used in its implementation. There is, as we have already stated, no fundamental reason for simulations of the coupled mode equations to be significantly slower than those based on the driven and damped NLS equation. The latter is usually solved by using standard tools developed for the simulation of the ordinary NLS equation in e.g. optical fibers. Numerical codes are commonly used that are based on the split-step Fourier method \cite{Agrawal}. The computationally expensive nonlinear step in simulations using coupled mode equations can, in a similar manner to the split-step method, be performed in the time-domain using the Fast Fourier Transform. This allows for a significant reduction in computational complexity, since the FFT algorithm is $\mathcal{O}(N \log N)$ instead of $\mathcal{O}(N^2)$ for a direct matrix calculation. To apply the FFT, we first rewrite the coupled mode equations in a suitably normalized form, viz.
\begin{align}
    & \frac{\partial \tilde{A}_\mu}{\partial \tilde{\tau}} = \left(-\frac{\kappa_\mu}{\kappa_0}+i\Omega_\mu\right)\tilde{A}_\mu + \delta_{\mu}f_0 + \nonumber\\
    & i\sum_{\alpha,\beta,\gamma}\delta_{\mu-(\alpha-\beta+\gamma)}\tilde{A}_{\alpha}\tilde{A}^*_{\beta}\tilde{A}_{\gamma},
    \label{eq:nCME}
\end{align}
where we have introduced: $\tilde{A}_\mu = \sqrt{2g_0/\kappa_0}A_\mu e^{i(\omega_\mu-\omega_{in}-\Delta\omega_{FSR})\tau}$, $\tilde{\tau} = (\kappa_0/2)\tau$, $\Omega_\mu = 2[(\omega_\mu-\omega_0)-(\omega_{in}-\omega_0)-\Delta\omega_{FSR}]/\kappa_0$ and $f_0 = \sqrt{8g_0\kappa_{ext}/\kappa_0^3}A_{in}$, cf. \cite{Kippenberg2}. The delta function in the nonlinear term stems from the energy conservation requirement that only those components that satisfy the relation $\omega_\mu = \omega_\alpha-\omega_\beta+\omega_\gamma$ should contribute to the field of mode $\mu$. Using tensor notation we may rewrite this term as
\begin{equation}
    \delta_{\mu,\alpha-\beta+\gamma}\tilde{A}_\alpha \tilde{A}^*_\beta \tilde{A}_\gamma = \tilde{A}_{\mu+\beta-\gamma} \tilde{A}^*_\beta \tilde{A}_\gamma = \tilde{A}_{\mu+k} \tilde{A}^*_{\gamma+k} \tilde{A}_\gamma,
\end{equation}
with the dummy index $k = \beta-\gamma$. The right hand side of this expression can be recognized as being two autocorrelations, which can be calculated using products of the field's discrete Fourier transform, viz.
\begin{equation}
    \tilde{A}_{\mu+k} \tilde{A}^*_{\gamma+k} \tilde{A}_\gamma = \mathcal{F}^{-1}\left[|\hat{a}_j|^2\hat{a}_j\right] = \frac{1}{N}\sum_{j=0}^{N-1}(|\hat{a}_j|^2\hat{a}_j)e^{i2\pi j\mu/N},
\end{equation}
with $\hat{a}_j = \mathcal{F}\left[\tilde{A}_\mu\right] = \sum_{\mu=0}^{N-1}\tilde{A}_\mu e^{-i2\pi j\mu/N}$ denoting the forward transform. The computation time for coupled mode simulations can thus be reduced substantially, if FFT algorithms are used to calculate these transforms. However, note that the Fourier transform is defined with respect to the mode number $\mu$ and not the actual resonant frequencies $\omega_\mu$ of the normalized Eq.(\ref{eq:nCME}), which implies that the transform will generally correspond to a distorted temporal profile of the field.

It is seen that the transformed four-wave-mixing term takes the same form as the Kerr nonlinearity in the NLS equation. Indeed, this should not be surprising, since the coupled mode equations are fundamentally nothing but a discrete Fourier expansion of a driven and damped NLS type equation, with some possible additional terms. The LLE model provides a single partial differential equation, which when discretized becomes a coupled system of ordinary differential equations. The LLE is a time-domain description, which models the field using a two time-scale approach, while the coupled mode equations provide a frequency domain description that does not make explicit use of the fast time-scale.

\begin{figure}[ht]
  \centering
  \includegraphics[width=\linewidth]{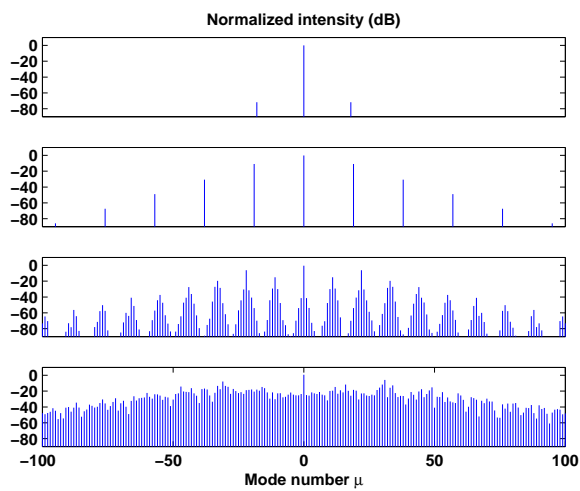}
  \caption{Examples of comb evolution for increasing pump intensity using coupled mode equations and split-step algorithm. Parameters identical to Fig. 10 of Ref. \cite{Chembo1}: $N = 256$, $\tilde{\tau} = 128\pi$, $\Omega_\mu = -0.00625\mu^2$, $\kappa_\mu = \kappa_0$ and $f_0 = \{1.01, 1.2, 1.8, 4\}\sqrt{2}$ in descending order. Total elapsed simulation time 54 seconds for the ensemble of four simulations.}
  \label{fig:combs}
\end{figure}

We will briefly mention two methods by which the coupled mode equations can be efficiently simulated. One may, e.g., use an ordinary ODE solver, such as a typical fourth-order Runge-Kutta solver, for the temporal evolution and compute the sum of the nonlinear contributions using FFTs. One may also use a split-step Fourier method of the same sort used to simulate the LLE, but consider the field to be in the frequency domain instead of the time-domain. The last method is perhaps most interesting since it blurs the difference between the solution of a system of exact coupled ODEs or a discretized PDE. We have implemented both methods and made comparisons of the simulation times for $N=201$ (zero padded to $N=2^8$) mode wide frequency combs with identical parameters to those considered by Chembo et al. in \cite{Chembo1}, who reported a time for a coupled mode simulation of a few days using a laptop computer. We found that either of the above methods allowed us to perform the same simulation in less than one minute using a standard desktop computer, see Fig. \ref{fig:combs}, which is on par with simulations based on the LLE equation, cf. \cite{Chembo2}.

Each different formalism has its own set of benefits and drawbacks. Coupled mode equations are often convenient to use, since they easily allow for frequency dependent absorption and coupling coefficients to be included. They can also be used to study cascaded frequency combs that have a non commensurate frequency spectrum, see \cite{Kippenberg2}. They are further convenient since they provide a direct frequency domain description of the optical comb. The drawback is that they do not give any direct time-domain information, so that if the temporal profile is desired it must be synthesized using the frequency components. It is also more difficult to extend the coupled system of equations in order to include the nonlinear effects of, e.g., self-steepening \cite{Gaeta} and Raman scattering \cite{Chembo2}. The driven and damped NLS equation has, conversely, previously been used to model dispersive fiber-ring cavities and it may be more familiar, since it is closely related to the ordinary NLS equation and its extensions for describing supercontinuum generation, which are widely used in nonlinear optics and other fields. The LLE model may also be more conductive to analytical developments, allowing for soliton solutions to be studied more easily, e.g., by means of perturbation theory \cite{Wabnitz}.

\section{Conclusions}

In conclusion, we have shown that numerical simulations using coupled mode equations can be as efficient as those based on the driven and damped nonlinear Schr\"odinger formalism. This is possible if the computationally expensive nonlinearity is computed in the time-domain, in a similar manner to the conventional split-step Fourier method. FFT algorithms can then be used, which provides a significant improvement in the computation time, resulting in orders of magnitude time savings over simulations reported by other methods in the literature. This permits for broadband combs to be readily simulated, and enables the modeling of octave spanning Kerr frequency combs consisting of not hundreds but thousands or more modes. Indeed, we have demonstrated that the driven and damped NLS and the coupled mode equations are essentially equivalent, as far as the computation time is involved, so that the choice of either formalism should not be based on simulation complexity. Rather, the choice of either approach should only depend on the particular convenience which is provided by the associated formalism for adequately modeling the problem at hand.

\section*{Acknowledgements}

This research was funded by Fondazione Cariplo, grant no. 2011-0395.


\begin{thebibliography}{99}
\bibitem{Kippenberg3}T. J. Kippenberg, R. Holzwarth and S. A. Diddams, Science 332 (2011) 555 % ``Microresonator-Based Optical Frequency Combs"
\bibitem{Matsko2}A. B. Matsko, A. A. Savchenkov, D. Strekalov, V. S. Ilchenko, and L. Maleki, Phys. Rev. A 71 (2005) 033804  % ``Optical hyperparametric oscillations in a whispering-gallery-mode resonator: Threshold and phase diffusion"
\bibitem{Chembo1}Y. K. Chembo, and N. Yu, Phys. Rev. A 82 (2010) 033801 % ``Model expansion approach to optical-frequency-comb generation with monolithic whispering-gallery-mode resonators"
\bibitem{Matsko1}A. B. Matsko, A. A. Savchenkov, W. Liang, V. S. Ilchenko, D. Seidel, and L. Maleki, Opt. Lett. 36 (2011) 2845 % ``Mode-locked Kerr Frequency Combs."
\bibitem{Hansson}T. Hansson, D. Modotto, and S. Wabnitz, submitted for publication in Phys. Rev. A (2013) % ``Dynamics of the Modulational Instability in Microresonator Frequency Combs"
\bibitem{Erkintalo2}S. Coen, H. G. Randle, T. Sylvestre, and M. Erkintalo, Opt. Lett. 38 (2013) 37  % ``Modelling of octave-spanning Kerr frequency combs using a generalized mean-field Lugiato-Lefever model"
\bibitem{Chembo2}Y. K. Chembo, and C. R. Menyuk, Phys. Rev. A 87 (2013) 053852 % ``Spatiotemporal Lugiato-Lefever formalism for Kerr-comb generation in whispering-gallery-mode resonators"
\bibitem{HTW}M. Haelterman, S. Trillo, and S. Wabnitz, Opt. Comm. 91 (1992) 401 % ``Dissipative modulational instability in a nonlinear dispersive ring cavity"
\bibitem{Gaeta}M. R. E. Lamont, Y. Okawachi, and A. L. Gaeta, arXiv:1305.4921 [physics.optics] (2013) % ``Route to stabilized ultrabroadband microresonator-based frequency combs"
\bibitem{Agrawal}G.~P.~Agrawal, Nonlinear Fiber Optics 4th ed, Academic Press, San Diego, 2007
\bibitem{Kippenberg2}T. Herr, K. Hartinger, J. Riemensberger, C. Y. Wang, E. Gavartin, R. Holzwarth, M. L. Gorodetsky, T. J. Kippenberg, Nature Phot. 6 (2012) 480 % ``Universal formation dynamics and noise of Kerr-frequency combs in microresonators"
\bibitem{Kippenberg1}T. Herr, V. Brasch, J. D. Jost, C. Y. Wang, N. M. Kondratiev, M. L. Gorodetsky, and T. J. Kippenberg, arXiv:1211.0733v2 [physics.optics] (2012) % ``Mode-locking in an optical microresonator via soliton formation"
\bibitem{Wabnitz}S. Wabnitz, J. Opt. Soc. Am. B 13 (1996) 2739 % ``Control of soliton train transmission, storage, and clock recovery by cw light injection"
\end{thebibliography}
\end{document}